\documentclass{aa} 
\usepackage{lscape}
\usepackage{graphicx}
\usepackage{txfonts}
\begin{document}
\title{The widest ultracool binary}
%
%
\author{Jos\'e A. Caballero\inst{1}\fnmsep\thanks{Alexander von Humboldt Fellow
at the MPIA.}} 
%
%
\institute{
Max-Planck-Institut f\"ur Astronomie, K\"onigstuhl 17, D-69117 Heidelberg,
Germany, \email{caballero@mpia.de}}
\date{Received 24 November 2006; accepted ... December 2006}

\abstract
{}   
{I test the ejection scenario of formation of brown dwarfs and very low-mass
stars through the detection of a very wide ultracool binary.}   
{LEHPM 494 (M6.0 $\pm$ 1.0 V) and DENIS-P J0021.0--4244 (M9.5 $\pm$ 0.5 V) are
separated by 1.3\,arcmin and are high proper motion co-moving ultracool stars. 
I have used six astrometric epochs spaced 22 years to confirm their common
tangential velocity.}   
{The angular separation between both low-mass stars keeps constant with an
uncertainty less than 0.1\,\%. 
I have also derived their most probable heliocentric distance (23 $\pm$ 2\,pc),
age interval (2--10\,Ga) and masses (0.103 $\pm$ 0.006 and 0.079 $\pm$
0.004\,M$_\odot$).  
The pair, with a projected physical separation of 1800 $\pm$ 170\,AU, is
by far the widest ultracool binary ever found in the field.}  
{This serendipitous and simple detection is inconsistent with ultra low-mass
formation ejection scenarios and complements current searches of low-mass
tight binaries.} 
\maketitle
%

\section{Introduction}
\label{intro}

\object{DENIS-P J0021.0--4244} (hereafter DE0021--42), with a spectral type
$\ge$ M9 V, was at the end of the previous decade one of the coolest known
isolated field dwarfs (Tinney et al. 1998; Delfosse et al. 1999).
At that time, only a few benchmark cooler objects had been identified as
companions to more massive bodies (Becklin \& Zuckerman 1988; Nakajima et al.
1995; Rebolo et al. 1998) or free floating (Ruiz et al. 1997; Kirkpatrick et al.
1997).     
Since then, hundreds of stars and brown dwarfs with very late M, L and T
spectral types have been discovered in direct imaging (see a full list at {\tt
DwarfArchives.org}). 
Many of them are faint companions to stars at separations between 15 and
$\sim$3600\,AU (with mass ratios $q \equiv \frac{\rm{M}_2}{\rm{M}_1} <$ 0.5;
Kirpatrick 2005, Burgasser, Kirkpatrick \& Lowrance 2005 and references
therein) or form tight binary systems with separations smaller than
$\sim$20\,AU (with mass ratios $q >$ 0.5; Bouy et al. 2005; Burgasser et
al. 2006).      
The masses and effective temperatures of late M-, L- and T-type objects are
below $\sim$0.15\,$M_\odot$ and $\sim$3000\,K, respectively, which leads to
classify them as ``ultracool'' objects.  

There are only three known {\em relatively wide} ultracool binary systems {\em
in the field} with separations of the order of 30--40\,AU and total masses below
$\sim$0.2\,$M_\odot$ (Harrington, Dahn \& Guetter 1974; Phan-Bao et al. 2005;
Burgasser \& McElwain 2006). 
Only very recently, a very wide low-mass binary, with a physical
separation significatively larger than the rest, has been discovered ($r
\approx$ 220\,AU, M$_1 \approx$ 0.090\,$M_\odot$, M$_2 \approx$
0.075\,$M_\odot$; Bill\`eres et al. 2005).  
Wide very low-mass binaries like this are unexpected in the embryo-ejection
scenario models, which also under-predict the observed frequency of tight
low-mass binaries (Bate et al. 2002; Sterzik \& Durisen 2003; Delgado Donate et
al. 2004; Umbreit et al. 2005). 

In young star forming regions (1--120\,Ma), some fragile systems with extremely
low binding energies (i.e. with ultra-low masses and/or very large separations)
have been reported (e.g. Mart\'{\i}n et al. 1998; Luhman 2004, 2005; L\'opez
Mart\'{\i} et al. 2004, 2005; Bouy et al. 2006; Jayawardhana \& Ivanov
2006).  
The least bound system known may be the \object{SE 70} + \object{S\,Ori 68}
pair in the 3\,Ma-old $\sigma$ Orionis cluster ($r \approx$ 1700\,AU, M$_1
\approx$ 0.045\,$M_\odot$, M$_2 \approx$ 0.005\,$M_\odot$; Caballero et al.
2006).   
On the contrary to systems in the field, the ``confirmation'' of
multiplicity in systems in young clusters does not come from proper-motion
measurements, but from statistical considerations on the separations to
background sources and to the rest of (candidate) cluster members. 
It is expected that most of these systems, maybe formed in the same way as low
mass stars and not in a protoplanetary disc, will not survive the tidal
disruption within the cluster. 
Nevertheless, they are still a problem to solve in the ejection scenario 
because it predicts that low-mass wide systems are torn apart at ages much less
than 3\,Ma. 

In this Letter I present the least bound binary system in the field with common 
proper-motion confirmation.

\section{Analysis and results}

   \begin{table}
      \caption[]{Data of Koenigstuhl 1 A and Koenigstuhl 1 B.}  
         \label{ko1ab}
     $$ 
         \begin{tabular}{lcccc}
            \hline
            \hline
            \noalign{\smallskip}
 				& K\"o\,1A		& K\"o\,1B		& Unit		& Ref.$^{a}$ \\  
            \noalign{\smallskip}
            \hline
            \noalign{\smallskip}
Name 				& LEHPM 494		& DE0021.0--42		& 		& 1,2 \\  
$\alpha$ (J2000)		& 00 21 10.42		& 00 21 05.74		& 		& 3 \\  
$\delta$ (J2000) 		& --42 45 40.0 		& --42 44 43.3		& 		& 3 \\  
$\mu_\alpha \cos{\delta}$	& +268$\pm$10		& +270$\pm$11		& mas\,a$^{-1}$	& 3 \\  
$\mu_\alpha \cos{\delta}$	& +246$\pm$15		& +250$\pm$9		& mas\,a$^{-1}$ & 4 \\  
$\mu_\delta$			& --21$\pm$8		& +4$\pm$10		& mas\,a$^{-1}$	& 3 \\  
$\mu_\delta$ 			& --48$\pm$18		& +18$\pm$15		& mas\,a$^{-1}$	& 4 \\  
$B_J$  				& 18.562		& 22.694		& mag		& 3 \\  
$R_1$  				& 16.033		& 19.633		& mag		& 3 \\  
$R_2$  				& 16.181		& 19.784		& mag		& 3 \\  
$I_N$  				& 13.599		& 16.697		& mag		& 5 \\  
$I$  				& 13.92$\pm$0.03	& 16.79$\pm$0.10	& mag		& 6 \\  
$J$  				& 12.00$\pm$0.02	& 13.52$\pm$0.02	& mag		& 6 \\  
$H$  				& 11.37$\pm$0.03	& 12.81$\pm$0.02	& mag		& 6 \\  
$K_{\rm s}$  			& 11.05$\pm$0.03	& 12.30$\pm$0.02	& mag		& 6 \\  
            \noalign{\smallskip}
            \hline
            \noalign{\smallskip}
Sp. type 			& M6.0 $\pm$ 1.0 V	& M9.5 $\pm$ 0.5 V	& 		& 7,8 \\  
$\log{\frac{L_{H\alpha}}{L_{\rm bol}}}$	& --		& --5.62		& 		& 7 \\  
$v \sin{i}$  			& --			& 17.5 $\pm$ 2.5	& km\,s$^{-1}$ 	& 7 \\  
$M_J$  				& 10.2 $\pm$ 0.2	& 11.7 $\pm$ 0.2	& mag 		& 8 \\  
Mass  				& 0.103 $\pm$ 0.006	& 0.079 $\pm$ 0.004	& M$_\odot$ 	& 8 \\  
T$_{\rm eff}$  			& 2850 $\pm$ 100	& 2250 $\pm$ 100	& K  		& 8 \\  
            \noalign{\smallskip}
            \hline
         \end{tabular}
     $$ 
\begin{list}{}{}
\item[$^{a}$] References:
1: Pokorny et al. (2003);
2: Tinney et al. (1998);
3: SSA (Hambly et al. 2001);
4: USNO-B1/NOMAD1 (Monet et al. 2003; Zacharias et al. 2004);
5: DENIS (Epchtein et al. 1997);
6: 2MASS (Cutri et al. 2003);
7: Basri et al. (2000), Mohanty \& Basri (2003);
8: this work.
\end{list}
   \end{table}
%

\begin{figure}
\centering
\includegraphics[width=0.49\textwidth]{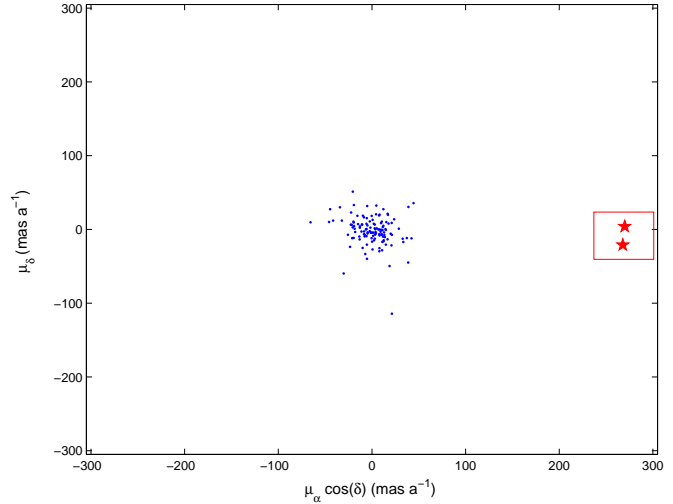}
\caption{Proper motions from SSA of all the sources in a 5-arcmin radius centred
in K\"o\,1A with errors less than 30\,mas\,a$^{-1}$.
Background sources are plotted with small dots.
K\"o\,1A and K\"o\,1B are shown with filled stars.
The box indicates the 2$\sigma$ uncertainty on the average proper motion of the
binary.}  
\label{propermotion}
\end{figure}

DE0021--42, which has been now classified as a normal M9.5V field dwarf 
(Basri et al. 2000; Mohanty \& Basri 2003), is located at $\sim$1.3\,arcmin to
the northwest of the faint high proper motion star \object{LEHPM 494} (see
finding chart in Figure \ref{RIH}).
The latter was discovered in the Liverpool-Edinburgh survey by Pokorny, Jones \&
Hambly (2003), and has gone unnoticed until now. 
Their only known physical parameters were the proper motion and photographic
$B_J$- and $I$-band magnitudes.
A rutinary visual inspection of the tangential velocities tabulated by
USNO-B1/NOMAD1 serendipitiously showed that the proper motions of DE0021--42 and
LEHPM 494 are extraordinary similar and very different from those of background
objects.  
USNO-B1/NOMAD1 proper-motion uncertainties are in general understimated (e.g.
Caballero et al. 2006), so I also studied the tangential velocities tabulated by
the SuperCOSMOS Science Archive (SSA). 
Figure \ref{propermotion} illustrates the high proper motion, apparently common,
of DE0021--42 and LEHPM 494.
In the upper part of Table \ref{ko1ab}, I provide the catalogued coordinates,
tangential velocities and magnitudes of both objects.
From now on, I will give to the pair the name \object{Koenigstuhl 1} (LEHPM 494
$\equiv$ Koenigstuhl 1 A, K\"o\,1A; DE0021--42 $\equiv$ Koenigstuhl 1 B,
K\"o\,1B; new wide low-mass binaries detected in an on-going survey will
have the same designation -- Caballero in prep.).   

%

\begin{figure}
\centering
\includegraphics[width=0.49\textwidth]{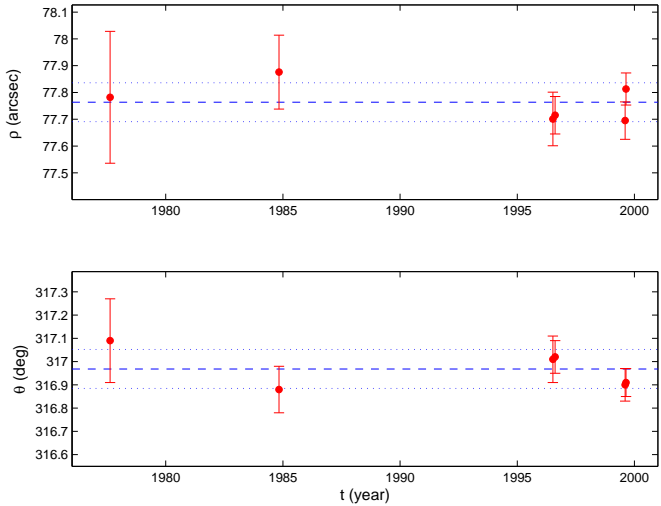}
\caption{Projected angular separations, $\rho$, and parallatic angles, $\theta$,
for the six epochs.
Baseline covers 22 years from 1977 to 1999.
Dashed and dotted lines indicate the average and $\pm 1 \sigma$ values,
respectively.}  
\label{rhotheta}
\end{figure}

I have used six imaging epochs, with a time baseline of 22 years, to measure the
variation of the angular separation, $\rho$, and orientation or parallactic
angle, $\theta$, between K\"o\,1A and K\"o\,1B. 
The used epochs were (in parenthesis, the passband(s) and the epoch of
observation):
UK Schmidt blue survey ($B_J$, J1977.626),
ESO Schmidt ($R_1$, J1984.878), 
DENIS ($IJK$, J1996.520),
UK Schmidt red survey ($R_2$, J1996.617),
UK Schmidt near-infrared survey ($I_N$, J1999.603) and
2MASS ($JHK_{\rm s}$, J1999.643).
I used the coordinates provided by the DENIS and 2MASS catalogues to compute
$\rho$ and $\theta$ at their respective epochs.
For the other epochs, I downloaded 10\,arcmin-wide fits images centred in K\"o\,1A
from the SuperCOSMOS Science Archive\footnote{The pixel size of the SuperCOSMOS
Sky Surveys is smaller than the Digital Sky Surveys DSS-I and DSS-II.},
and performed standard astrometric measurements within IRAF.
I used relatively bright USNO-B1 sources in the field of view with null proper
motion to compute the scale and orientation of the plates. 
In Figure \ref{rhotheta}, I show the variation of $\rho$ and $\theta$ with
time.
They maintain constant with uncertainties as low as 0.09\,\% (see Table
\ref{KO1AB}).   
The average projected angular separation between K\"o\,1A and K\"o\,1B is $\rho$
= 1.296 $\pm$ 0.012\,arcmin.  
The modulus of the mean proper motion of the system, $\mu$ =
258\,mas\,a$^{-1}$, is 14 times larger than the median of the tangential
velocities of the background sources tabulated by SSA in a region of a radius of
5\,arcmin centred in K\"o\,1A.  
After 22 years, the pair has travelled 5.7\,arcsec, which is $\sim$80 times
larger than the uncertainty in the average $\rho$. 
The probability of two field objects being separated $\sim$1.3\,arcmin, sharing
the same tangential velocity and {\em not} being gravitationally bound is
insignificant.
Therefore, K\"o\,1A and K\"o\,1B are doubtless a common proper motion pair.

%

\begin{figure}
\centering
\includegraphics[width=0.49\textwidth]{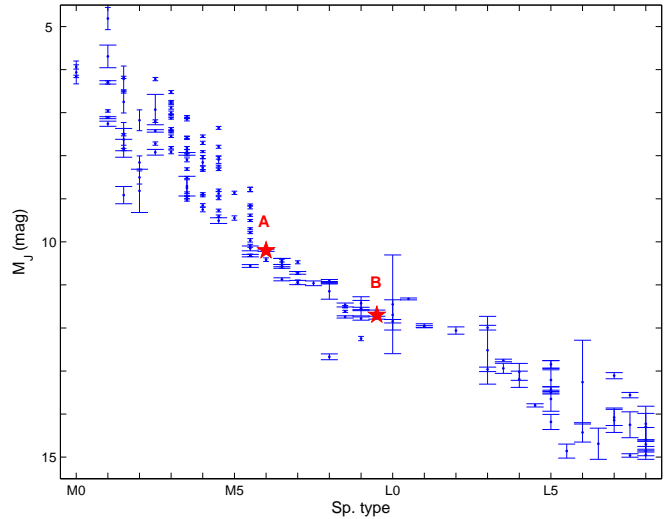}
\caption{$M_J$ vs. spectral type diagram.
Ultracool dwarfs with parallax determination and $J$-band magnitude from 2MASS
are shown with small dots with errorbars.
Expected positions of K\"o\,1A and K\"o\,1B for their most probables $M_J$ and
spectral types (measured from optical spectra in the case of K\"o\,1B) are
indicated with filled stars.
The cubic polynomial fit is not shown for clarity.}  
\label{HR}
\end{figure}

I have collected an extensive sample of field M- and L-type dwarfs with spectral
type and parallax determinations and homogeneous 2MASS photometry.
The distance measurements come from Perryman et al. (1997) (in the case of
companions to {\em Hipparcos} stars), Dahn et al. (2002), Vrba et al. (2004) and
the List of the Nearest 100 Stellar Systems of the Research Consortium on Nearby
Stars ({\tt www.chara.gsu.edu/RECONS}; maintained by T. J. Henry).
I have fitted to a cubic polynomial the relation between absolute $J$-band
magnitude, $M_J$ = $J + 5 - 5 \log{d}$, and spectral type in the interval
M0--L9 (Figure \ref{HR}).
I have computed the heliocentric distance to K\"o\,1B at 23 $\pm$ 2\,pc from the
fit using its $J$-band magnitude and its spectral type (M9.5 $\pm$ 0.5 V; Basri
et al. 2000). 
I have also derived the $M_J$ and the most probable spectral type of K\"o\,1A
(M6.0 $\pm$ 1.0 V) from its $J$-band magnitude and the distance deduced for
K\"o\,1B. 
Uncertainties come from the error in the fit.  
K\"o\,1A displays colours from SSA, DENIS and 2MASS photometry that are
very similar to other standard M5.5--6.5 field dwarfs (\object{Proxima
Centauri}, \object{DX Cancri}), which supports the derived spectral type. 
The same occurs to K\"o\,1B when compared to other M9.0--9.5 field dwarfs
(\object{GJ 3517}, \object{DY Piscium}).
The projected physical separation of Koenigstuhl 1 is 1800 $\pm$ 170\,AU,
the widest found among ultracool binaries.   

Through comparison of $M_J$ to theoretical models of the Lyon group (Baraffe et
al. 1998; Chabrier et al. 2000), I have estimated the masses of K\"o\,1A and
K\"o\,1B at 0.103 $\pm$ 0.006 and 0.079 $\pm$ 0.004\,M$_\odot$. 
Both NextGen98 and Cond00 models provide similar results.
The total mass and mass ratio of the system are 0.182 $\pm$ 0.007\,M$_\odot$ and
$q$ = 0.77 $\pm$ 0.06, comparable to those derived for tight
low-mass binary systems.  
The long orbital period, of $\sim$0.2\,Ma, will prevent any astrometric mass
measurement.
I have used a most probable age of the system in the interval between 2 and
10\,Ga for several reasons:
($i$) the absence of Li {\sc i} $\lambda$6707.8\,\AA~in absorption in the
optical spectrum of K\"o\,1B (M9.5-type stars or brown dwarfs younger than
$\sim$1\,Ga are expected to display lithium in absorption);
($ii$) the low activity of K\"o\,1B based on its H$\alpha$ emission
(pEW(H$\alpha$) = +0.5\,\AA; Mohanty \& Basri 2003); and
($iii$) Koenigstuhl 1 does not share the motion of any young stellar kinematic
group (the Galactic space-velocity components UVW have been computed from the
average tangential velocity and the radial velocity of K\"o\,1B from Mohanty \&
Basri 2003, V$_r$ = +2 $\pm$ 1\,km\,s$^{-1}$). 
If younger than 1\,Ga, then K\"o\,1B would be a brown dwarf.
Tables \ref{ko1ab} and \ref{KO1AB} summarise the data of the binary system and
their components. 
The gravitational potential energy of K\"o\,1AB, $U_{g}$, is similar to
those of the very young low-mass systems found by Chauvin et al. (2004), Luhman
(2004) and Jayawardhana \& Ivanov (2006).

Basri \& Reiners (2006) recently claimed a marginal (3 $\sigma$) detection of
spectroscopic binarity in K\"o\,1B, which would make Koenigstuhl 1 to be a
hierarchical triple system. 
In the case of K\"o\,1B being an equal-mass binary, then Koenigstuhl 1 ABab would
be located at about 33\,pc and be formed by an 0.11\,M$_\odot$-mass primary
separated 2500\,AU of a 0.08 + 0.08\,M$_\odot$-mass binary.
The hypothetical ABab triple system would have roughly the same binding energy
as the simple AB binary ($U_{g}$ = --1.2 10$^{33}$\,J).
If confirmed, K\"o\,1ABab would be a low-mass analog to the system
\object{G 124--62} ABab (M$_{\rm total} \approx$ 0.73\,M$_\odot$, $r \approx$
1500\,AU; Seifahrt, Guenther \& Neuh\"auser 2005).
Wide triples like K\"o\,1ABab, G 124--62 ABab, \object{GJ 1001} ABab,
\object{$\epsilon$ Ind} ABab and \object{GJ 417} ABab may be prevalent in
nature over wide binaries.

\section{Conclusions}

Assuming K\"o\,1B is a single object, then Koenigstuhl 1 AB is eight times wider
than the Bill\`eres et al. (2005) system, about 50 times wider than other
relatively wide ultra low-mass binaries in the field, and three orders of
magnitude wider than the widest ``normal'' tight ultracool binary.  
Besides, it is 40 times wider than the maximum projected separation of low-mass
binaries scaled with the total mass found by Burgasser et al. (2003). 
Koenigstuhl 1 AB is also much wider than the binaries found in young clusters.
Its projected physical separation is only comparable to those of the \object{Hn
12} A + B (L\'opez Mart\'{\i} et al. 2005) and SE 70 + S\,Ori 68 (Caballero et
al. 2006) very young systems.  
The separation could be even larger if K\"o\,1B were a tight binary. 

   \begin{table}
      \caption[]{Data of the Koenigstuhl 1 AB binary system.}  
         \label{KO1AB}
     $$ 
         \begin{tabular}{lcc}
            \hline
            \hline
            \noalign{\smallskip}
Quantity 	 		& Value 	& Unit  	\\  
            \noalign{\smallskip}
            \hline
            \noalign{\smallskip}
$\mu_\alpha \cos{\delta}$ 	& +258$\pm$12$^a$ & mas\,a$^{-1}$ \\  
$\mu_\delta$ 			& --12$\pm$29$^a$ & mas\,a$^{-1}$ \\  
$\rho$  			& 77.76 $\pm$ 0.07& arcsec	\\  
$\theta$  			& 316.97 $\pm$ 0.08& deg	\\  
$d$  				& 23 $\pm$ 2	& pc 		\\  
$r$ 				& 1800 $\pm$ 170& AU 		\\
$U_{g}$ 			& --8.0 $\pm$ 1.0 & $10^{33}$\,J \\
$P$  				& $\sim$2 10$^5$ & a 		\\  
$U$  				& --23 $\pm$ 3	& km\,s$^{-1}$ 	\\  
$V$  				& --15 $\pm$ 3	& km\,s$^{-1}$ 	\\  
$W$  				& --4.9 $\pm$ 1.3& km\,s$^{-1}$ \\  
Age 				& 2--10		& Ga 		\\
            \noalign{\smallskip}
            \hline
         \end{tabular}
     $$ 
\begin{list}{}{}
\item[$^{a}$] Average of the tangential velocities of K\"o\,1A and K\"o\,1B
tabulated by SSA and USNO-B1/NOMAD1.
\end{list}
   \end{table}

There are already in the literature excellent discussions on how wide low-mass
binaries confront current hydrodynamical and N-body simulations of the
fragmentation in molecular clouds
(see references in Section \ref{intro}, especially Bill\`eres et al. 2005). 
This confrontation is even harder to explain in the case of the relatively old
Koenigstuhl 1 AB system, which is by far the least bound binary in the field.
Very probably, there are similar systems, long awaiting to be discovered.
Were their original physical separations so wide? 
Or have they been perturbed by encounters with massive objects as they travel in
the Galaxy?
The answer may come from the measurement of the variation of the frequency of
wide ultracool binaries in the solar neighbourhood as a function of the
separation from tens to thousands AU.
This project is achievable with already-available astronomical databases and is
complementary to the study of tight low-mass binaries, which in contrast
requires expensive high spatial resolution imaging facilities.  
If Galactic perturbation has not played an important r\^ole, then Koenigstuhl 1
AB should be seriously taken into account for further low-mass star-forming
scenarios.

\begin{acknowledgements}

I thank A. Burgasser for his fast and efficient refereeing.
I also thank B. Goldman, J. Maldonado and R. Mundt for providing me
helpful comments and UVW data.
I have used IRAF, the M, L, and T dwarf compendium housed at {\tt
DwarfArchives.org}, the List of the Nearest 100 Stellar Systems, the Two-Micron
All Sky Survey, the Deep Near Infrared Survey of the Southern Sky, the USNO-B1
and NOMAD catalogues, the SuperCOSMOS Science Archive and the SIMBAD database.
Partial financial support was provided by the Spanish Ministerio de
Ciencia y Tecnolog\'{\i}a project AYA2004--00253 of the Plan Nacional de
Astronom\'{\i}a y Astrof\'{\i}sica. 
 
\end{acknowledgements}

\appendix

\section{Finding chart}

\begin{figure}
\centering
\includegraphics[width=0.49\textwidth]{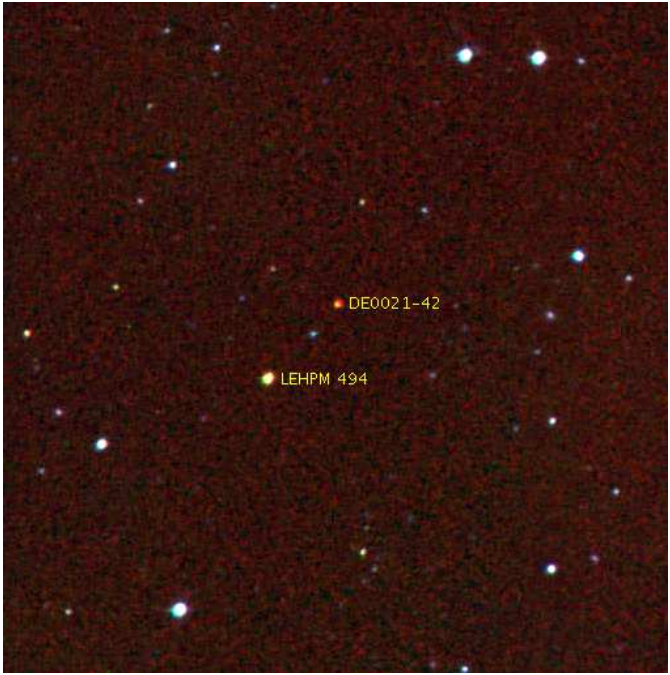}
\caption{False-colour finding chart, 10 $\times$ 10\,arcmin$^2$ wide, showing
the Koenigstuhl 1 binary system.
Both components are labelled (LEHPM 494 $\equiv$ K\"o\,1A, DE0021--42 $\equiv$
K\"o\,1B). 
Red is for $H$, green is for $I$ and blue is for $R$.} 
\label{RIH}
\end{figure}

{\em Note to the Editor: Figure \ref{RIH} is available only in the electronic
version of the journal.}

\end{document}